\documentclass[prl,aps,a4paper,twocolumn,twoside,showpacs,preprintnumbers,amsfonts,amsmath,amssymb,floatfix]{revtex4-1}

\usepackage{graphicx}
\begin{document}

\title{Fluctuation-preserving coarse graining for biochemical systems}
\newcommand{\eq}[1]{Eq.~(\ref{#1})}
\newcommand{\ie}{\emph{i.e., }}
\newcommand{\cf}{\emph{cf. }}
\newcommand{\eg}{\emph{e.g. }}

\author{Bernhard Altaner}
\author{J\"urgen Vollmer}
\affiliation{Max-Planck-Institut f\"ur Dynamik und Selbstorganisation (MPIDS), 37077 G\"ottingen, Germany}
\affiliation{Fakult\"at f\"ur Physik, Universit\"at G\"ottingen, 37077 G\"ottingen, Germany}

\date{\today}

\begin{abstract}
  Finite stochastic Markov models play a major role in modeling biological systems.
  Such models are a coarse-grained description of the underlying microscopic dynamics and can be considered {\it mesoscopic}.
  The level of coarse-graining is to a certain extend arbitrary since it depends on the resolution of accomodating measurements.
  Here, we present a systematic way to simplify such stochastic descriptions which preserves both the meso-micro and the meso-macro connection.
  The former is achieved by demanding locality, the latter by considering cycles on the network of states.
  Our method preserves fluctuations of observables much better than na\"ive approaches.
\end{abstract}

\maketitle
In recent years non-equilibrium fluctuations have become the center interest of {\it stochastic thermodynamics} \cite{Esposito+vdBroeck2010,Liepelt+Lipowsky2007}.
Rare events in situations far from equilibrium can now be universally described by fluctuation theorems \cite{Seifert2005,Seifert2005epl,Faggionato+dPietro2011}.
Intensive stochastic modelling of biophysical processes has started in the 1960s with Hill's cycle kinetics \cite{Hill1966,Hill1977} (where the focus lies on averages) and is still a very active field of research, though attention has shifted to the importance of fluctuations, \cf Ref. \cite{Hallatschek2010}.

Although Hill's methods were designed for biological problems, they have lead to general insights in statistical physics \cite{Schnakenberg1976} and mathematics \cite{Kalpazidou2006,Jiang_etal2004}.
It was understood that in non-equilibrium situations currents driven by non-trivial forces which are usually called affinities.
Assigning these affinities to cycles on the network of states rather then to the states themselves, they have a direct thermodynamic interpretation \cite{Schnakenberg1976,Faggionato+dPietro2011}.
This hints at possible redundancy in the description and already Hill asked how and when a network reduction would be possible.
In statistical physics, such reduction are often summarized under the term of {\it coarse-graining} (CG) methods.
It was recently shown for a special CG procedure that the ability to capture fluctations depends on the preservation of cycle topology of the network \cite{Puglisi_etal2010}.
%
%

In this Letter we present a new paradigm for coarse-graining of stochastic dynamics which preserves the non-equilibrium steady-state fluctuations of physical currents.
Though we focus on biological situations the method can be universally applied to any finite model of stochastic thermodynamics.
Our method is based on two requirements:
 {\bf (i)} The preservation of the {\it algebraic} and {\it topological} structure of the {\it cycles} of the network and {\bf (ii)} {\it locality}.
 Further, {\bf (iii)} the variation of the system's entropy along single trajectories \cite{Seifert2005} is considered to close the equations.
\begin{figure}[h]
  \begin{center}
      \includegraphics[width=.48 \textwidth]{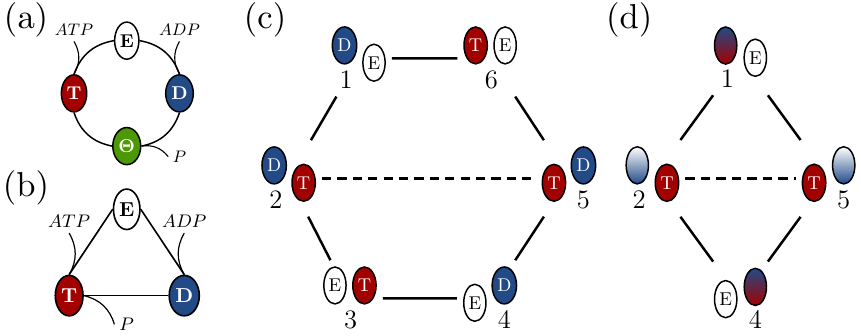}
      \vspace{-9mm}
  \end{center}
  \vspace{-2mm}
  \caption{(color online) (a) The catalytic cycle at kinesin's active site. In a four-stage process $A\boldsymbol{T}P$ binds to the {\bf E}mpty molecule and is split into $\boldsymbol{\Theta} = ADP+P$. Then first $P$ and later $A\boldsymbol{D}P$ is released. Since the release of $P$ happens immediately after the splitting, often a three-stage process (b) is assumed where the state $\Theta$ is absorbed into its neighbor states.
  (c) 6-state model of kinesin \cite{Liepelt+Lipowsky2007}.
  The dashed line is the mechanical transition that allows the motor to move.
  (d) Coarse-grained description with states $3$ and $6$ reduced.}
  \label{fig:reductionkinesin}
  \vspace{-2mm}
\end{figure}
To illustrate our method we consider the molecular motor kinesin which is able to perform directed motion along intracellular filaments called microtubuli \cite{Yildiz_etal2004, Carter+Cross2005, Liepelt+Lipowsky2007,Lipowsky+Liepelt2008}.
It has two heads (active sites) where adenosin triphosphate ($ATP$) is catalytically split into adenosin diphosphate ($ADP$) and inorganic phosphate ($P$). 
During the reaction, the molecule undergoes a conformational change that couples the two active sites and induces a mechanical transition.
This allows the motor to ``walk'' in a ``hand-over-hand'' mechanism \cite{Yildiz_etal2004}.

\begin{figure}[t]
  \begin{center}
    \begin{minipage}[t]{.02\textwidth}%
      \vspace{0mm}
      (a)
    \end{minipage}
    \begin{minipage}[t]{.21\textwidth}%
      \vspace{0mm}%
       \includegraphics[height=.4\textwidth]{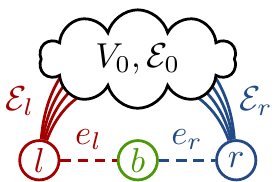}\\
       \includegraphics[height=.4\textwidth]{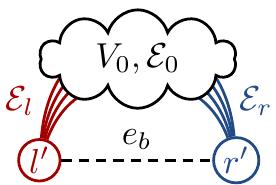}\vspace{-3mm}
    \end{minipage}
    \begin{minipage}[t]{.02\textwidth}%
      \vspace{0mm}
      (b)
    \end{minipage}
    \begin{minipage}[t]{.21\textwidth}%
      \vspace{0mm}
    \includegraphics[height=.42\textwidth]{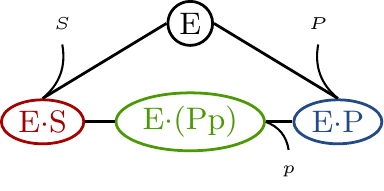}\\
    \includegraphics[height=.42\textwidth]{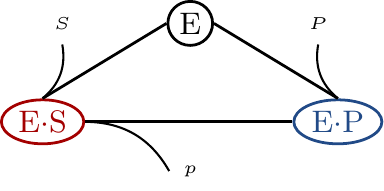}\vspace{-3mm}
    \end{minipage}
  \end{center}
  \vspace{-4mm}
  \caption{(color online) (a) Illustration of the coarse-graining procedure that leaves cycle topology constant:
  Reduction of a ``bridge'' $b$ will absorb the two dashed edges of the original graph (top) into one edge in the coarse-grained graph (bottom).
  This also leads to a change of rates along the edges $\mathcal E_l$ and $\mathcal E_r$ whereas edges $\mathcal E_0$ connecting only unchanged vertices $V_0$ remain unchanged.
  (b) Enzyme catalysis. An enzyme $E$ binds a substrate $S$ to form a complex $E\cdot S$. The substrate is split to form products $P$ and $p$ where the latter is always released first. The dynamics can be modeled with four (top) and three (bottom) states.} \label{fig:bridge-cg}
\end{figure}

The catalytic cycle of a single head (Fig.~\ref{fig:reductionkinesin}a) is an example of a general enzymatic activity (Fig.~\ref{fig:bridge-cg}).
This mesoscopic, stochastic description with its fluctuations has its origins in a microscopic, deterministically chaotic dynamics.
Here, we investigate how a stochastic description can be further simplified while preserving its fluctuations.

\paragraph{Stochastic Formalism}
We consider a Markov process on a finite number of mesoscopic states $i \in [1..N]$.
We call them {\it mesoscopic}, because for physical systems they amount to a partition of the underlying microscopic phase space.
Transitions between the states $i$ and $j$ occur with time-independent rate constants $w^i_j \geq 0$.
For simplicity, we assume that there is only one mechanism by which the transition between two states can happen (although a generalization is possible \cite{Esposito+vdBroeck2010,Faggionato+dPietro2011}).
Because of the reversibility of microscopic physical laws we demand {\it dynamical reversibility}, \ie $w^i_j>0\Leftrightarrow w^j_i>0$.
One can visualize the system as a graph $G=(V,\mathcal E)$ with the mesoscopic states as vertices $V$ and edges $\mathcal E$ where $w^i_j>0$.
At time $t$ the system will be in state $i$ with a probability $p_i(t)$.
The flux from state $i$ to state $j$ is
\begin{equation}
  \phi^i_j := p_iw^i_j,~i\neq j.
  \label{eq:flux-defn}
\end{equation}
Assuming connectedness of the network, a unique invariant distribution $p_i$ exists \cite{Feller1968}.
In the steady state the net influx to each state equals the net outflux (Kirchhoff's law for currents $I^i_j := \phi^i_j - \phi^j_i$),
\begin{equation}
  \sum_j I^j_i = \sum_j\left[ \phi^j_i -\phi^i_j \right]=\sum_j\left[ p_j w^j_i -p_i w^i_j \right]=0~\forall i.
  \label{eq:master-equation}
\end{equation}
The steady-state probability distribution can be calculated explicitly as a polynomial in the rates $w^i_j$ using a graph-theoretic matrix-tree method \cite{Hill1977}.
From now on, all variables are time-independent steady-state quantities unless mentioned otherwise.
Eq.~(\ref{eq:master-equation}) can be used to decompose the steady fluxes using different sets of cycles on $G$ \cite{Hill1977,Schnakenberg1976,Faggionato+dPietro2011}.
A cycle \(\alpha\) of length \(s_{\alpha}\) is an ordered set of vertices  which form a self-avoiding closed path, where we identify cycles differing only by a cyclical permutation of vertices.
In the following, when referring to cycles, we mean non-trivial cycles with $s_\alpha \geq 3$.
Central quantities for this work are the edge affinities $A^i_j = \log\left( \phi^i_j / \phi^j_i \right)$.
Along a cycle $\alpha= (i_1, i_2,\ldots,i_{s_{\alpha}})$ they add up to cycle affinities
\begin{equation}
  A_\alpha = 
  \sum_{k=1}^{s_\alpha}\log\left[ \frac{\phi^{i_{k-1}}_{i_{k}}}{\phi^{i_{k}}_{i_{k-1}}} \right] =
  \sum_{k=1}^{s_\alpha}\log\left[ \frac{w^{i_{k-1}}_{i_{k}}}{w^{i_{k}}_{i_{k-1}}} \right].
  \label{eq:cycle-affinity-defn}
\end{equation}
In physical models they take only values that reflect the macroscopic thermodynamic affinities \cite{Hill1977,Schnakenberg1976}.

\paragraph{Coarse graining}
We suggest a coarse-graining procedure based on natural requirements:\\
\hspace{2em} {\bf (ia)} {\it Cycle topology:} The number and mutual connections of cycles are preserved. This determines possible targets for the reduction.\\
\hspace{2em} {\bf (ib)} {\it Cycle affinities:} The algebraic values of the affinity of any cycle is preserved. This yields the connection to the macroscopic level, \ie thermodynamics.\\
\hspace{2em} {\bf (ii)} {\it Locality}: Fluxes, probabilities and observables may only change locally. This yields the connection to the microscopic level, \ie the microscopic phase space. \\
\hspace{2em} {\bf (iii)} {\it Trajectories:} The system's entropy variation along trajectories is preserved. This is a natural choice and closes the equations.

%
To demonstrate our method we address cycles that contain {\it bridge states}, like states $3$ and $6$ in Fig.~\ref{fig:reductionkinesin}. 
Bridges are connected to exactly two neighbor states that are themselves not connected to each other as shown in Fig.~\ref{fig:bridge-cg}a.
We use the index $b$ for the target bridge state, and $l$ or $r$ for the left or right neighbor.
Without loss of generality we assume that there is a positive net current $I = I^l_b = I^b_r>0$ flowing through the bridge from the left to the right neighbor state.
The other states, which must not be influenced by the procedure (\cf (ii)) are summarized in the set $V_0\subset V$.
In the CG procedure we absorb the bridge into its neighbors leading to new states $l'$ and $r'$ and adjust the transition rates for the sets of edges $\mathcal E_l$ and $\mathcal E_r$ connecting $l$ and $r$ to the rest of the network.
This has to be done in accordance with requirement (i)b yielding $A_\alpha = A'_\alpha$ for any cycle in the network.
Demanding the conservation of fluxes along any edge $e \in \mathcal E / \left\{ e_l,e_r \right\}$ not belonging to the bridge, Eq.~(\ref{eq:cycle-affinity-defn}) yields
\begin{subequations}%
\begin{equation}%
  {\phi^{l'}_{r'}}/{\phi^{r'}_{l'}}\stackrel!= \left(\phi^l_b \phi^b_r\right)/\left(\phi^r_b \phi^b_l\right). 
  \label{eq:CGflux}
\end{equation}%
Any trajectory passing through the two edges $(l,b)$ and $(b,r)$ in the original model will be a trajectory through $(l',r')$ in the coarse-grained model.
The change of a trajectory's entropy (starting from a steady-state ensemble) is the difference of the logarithms of the invariant distribution \cite{Seifert2005}.
With that, (iii) leads to
\begin{equation}%
  {p_{l'}}/{p_{r'}}\stackrel{!}{=} {p_l}/{p_r}.
  \label{eq:CGprob}
\end{equation}%
\label{eq:CG}%
\end{subequations}%
A priori, other closures of the form $p_{l'}/p_{r'} = c$ are also possible but lack the advantage of the stochastic thermodynamic interpretation.

Together with the steady-state balance condition (\ref{eq:master-equation}) and the locality assumption, the Eqs.~(\ref{eq:CG}) uniquely determine all rate constants of the coarse-grained model.
They can be found to be
\begin{subequations}%
  \begin{eqnarray}%
    w^i_{n'} &=& w^i_n\quad \mathrm{for}~i \in V_0, n \in \left\{ l,r \right\},\\
    w^{n'}_i &=& w^n_i/f\quad \mathrm{for}~i \in V_0, n \in \left\{ l,r \right\},\\
    w^{l'}_{r'} &=& (I+m)/(f p_l),\\ 
    w^{r'}_{l'} &=& (m)/(f p_r),
  \end{eqnarray}%
\label{eq:newrates}%
\end{subequations}%
where
\begin{subequations}%
  \begin{eqnarray}%
    f = (p_l + p_r + p_b)/(p_l + p_r),\\
    m = \phi^r_b \phi^b_l/(I + \phi^r_b + \phi^b_l).
  \end{eqnarray}%
  \label{eq:mf_defn}%
\end{subequations}%
One may argue that the method might not be practical, because one has to solve the original model to compute the correct rates for the simpler model.
However, the situation is better than this:
One only needs the steady state of the original model, which is accesible numerically to arbitrary precision by different means.
In particular, if the rates of the original model are subject to inaccurities either by measurement or by modeling, the numerical error (if there is any) to the steady-state probabilities is neglegible.

\paragraph{Single cycle: Simple catalysis model}
The easiest reducible topology is a cycle consisting of four states,\eg the enzyme catalysis presented in Figs.~\ref{fig:reductionkinesin}a and \ref{fig:bridge-cg}b where a transient intermediate state is identified as the bridge.
A na\"ive approximation for the new rates would be $w^{l'}_{r'}=w^l_b w^b_r \langle \tau_b \rangle$ and $w^{r'}_{l'}=w^r_b w^b_l \langle \tau_b \rangle$ where $\langle \tau_b \rangle^{-1} = w^b_l + w^b_r$ is the time constant for decay out of the bridge state.
Hill \cite{Hill1977} derived this result for three linearly connected states with the center one being transient.
This choice is also the basis of the method proposed in Ref.~\cite{Puglisi_etal2010}, where its shortcomings have already been discussed.
It fulfills
\begin{equation}
  {w^{l'}_{r'}}/{w^{r'}_{l'}} = \left(w^l_b w^b_r\right)/\left(w^r_b w^b_l\right).
  \label{eq:affinities-LL-choice}
\end{equation}
In Ref.~\cite{Lipowsky+Liepelt2008} it is interpreted as a condition on the local energy landscape and was used to reduce the enzymatic reaction of kinesin's active site (Fig.~\ref{fig:reductionkinesin}a). 
If all other rates are unchanged, Eq. (\ref{eq:affinities-LL-choice} preserves the affinity (\ref{eq:cycle-affinity-defn}) of the cycle.
eowever, in general, it leads to a non-local redistribution of steady-state probabilities.
Hence, it does not comply with our method.
This yields another motivation of Eq.~(\ref{eq:CGprob}).
One can easily check that this condition on the ratio of the new probabilities is the only one that leads to rates that fulfill Eq.~(\ref{eq:affinities-LL-choice}).
\paragraph{Fluctuations of physical observables}
\begin{figure}[t]
  \begin{center}
    \includegraphics{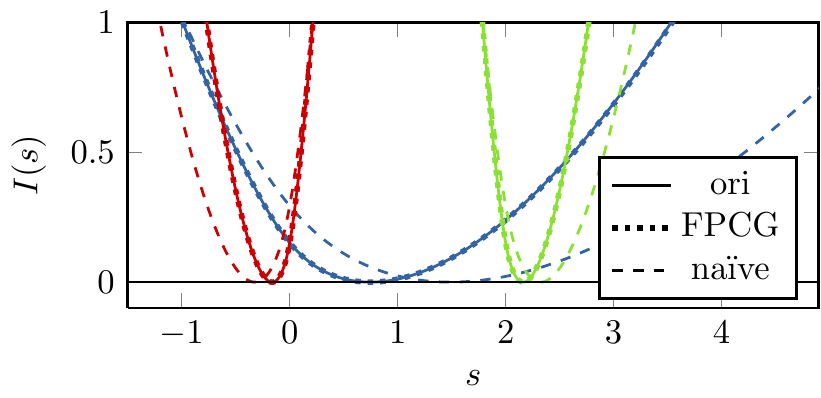}
  \end{center}
  \vspace{-8mm}
\caption{(color online) Large-deviation function (LDF) $I(s)$ for the entropy production (blue, center), the product (red, left) and the substrate (green, shifted to the right by $s_{0}=2$) association for the enzyme model (Fig.\,\ref{fig:bridge-cg}b).
  All transition rates are unity but the release rate for the first product $p$ which has the value $100$.
  The LDF obtained from the  fluctuation preserving coarse-graining method (``FPCG'') overlaps almost perfectly with the original model (``ori'') while the ``na\"ive'' choice strongly changes fluctuations.
}
  \vspace{-2mm}
  \label{fig:4stateFluctuations}
\end{figure}
Since the CG procedure changes the mesoscopic state space, also coarse-grained physical observables need to be defined.
We consider physical currents, which are modeled by anti-symmetric matrices \mbox{$O \in\mathbb R ^{N\times N}$} that assign a value $O^i_j = -O^j_i$ to each transition \mbox{$i\to j$}.
The observable $\widetilde O$ for the case where the bridge state has been eliminated has entries 
\begin{subequations}%
  \begin{eqnarray}%
    \widetilde O^{l'}_{r'} &=& O^l_b + O^b_r + (d_l - d_r),\\ 
    \widetilde O^{n'}_i &=&O^n_i + d_n \quad \mathrm{for}~i \in V_0, n \in \left\{ l,r \right\},\\
    \widetilde O^i_j &=& O^i_j\quad\mathrm{for}~i,j\in V_0.
  \end{eqnarray}%
  \label{eq:newCurrentObserable}%
\end{subequations}%
The constants $d_l$ and $d_r$ depend on the microscopic dynamics and the chosen partitioning of phase space.
As we do not know these details, $d_l$ and $d_r$ act as gauges that do not change the macroscopic observations.
Hence, in general, we choose $d_l \equiv d_r \equiv 0$ for simplicity.

A special observable, where the gauge is prescribed differently, is the quantity
\begin{equation}
  B^i_j = \log\left( {w^i_j}/{w^j_i} \right).
  \label{eq:basicFED}
\end{equation}
It is determined solely by the mesoscopic transition rates and therefore takes a special role.
Hill calls it the {basic free energy difference} between two mesoscopic states \cite{Hill1977}.
Recently, Seifert made this point more clear by stating the assumptions, under which it can be identified with the heat dissipated in the medium for transition $i \to j$ \cite{Seifert2005,Seifert2011}.
In an electric analogy it would be the electromotance \cite{Altaner_etal2012}.

One finds
\begin{subequations}%
  \begin{eqnarray}%
    \widetilde B^{l'}_{r'} &=& B^l_b + B^b_r, \label{eq:Bbridge}\\ 
    \widetilde B^{n'}_i &=& B^n_i - \log f \quad \mathrm{for}~i \in V_0, n \in \left\{ l,r \right\}, \label{eq:Bneighbour}\\
    \widetilde B^i_j &=& B^i_j\quad\mathrm{for}~i,j\in V_0. \label{eq:Brest}
  \end{eqnarray}%
  \label{eq:newElectromotance}%
\end{subequations}%
Eq.~(\ref{eq:Bbridge}) is the logarithm of Eq.~(\ref{eq:affinities-LL-choice}).
Eq.~(\ref{eq:Bneighbour}) states, that along the edges $\mathcal E_n,~n \in \left\{ l,r \right\}$, there is an additional contribution $- \log f$ to $\widetilde B^n_i$, which is the same for both neighbors due to the closure (\ref{eq:CGprob}).
Eq.~(\ref{eq:Brest}) expresses locality and is independent of the closure.
We note, that non-current observables defined on the states rather than the transitions can also consistently be transformed \cite{Altaner+Vollmer2012long}.

\begin{figure*}[t]
  \includegraphics{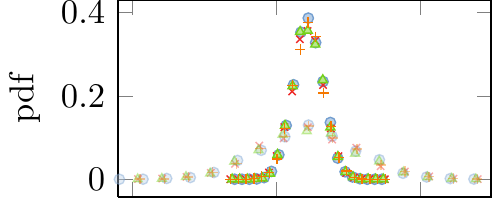}%
  \includegraphics{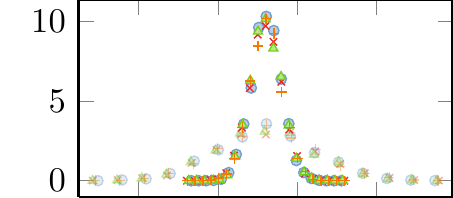}%
  \includegraphics{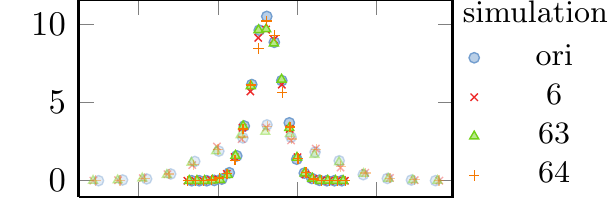}\\%
  \includegraphics{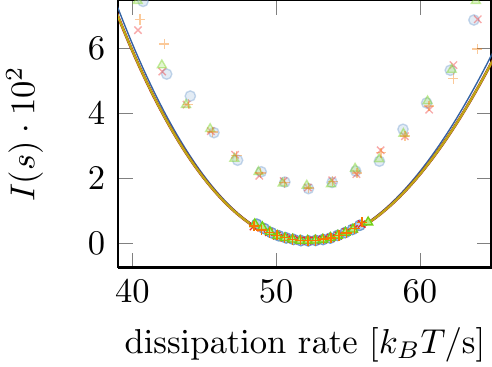}%
  \includegraphics{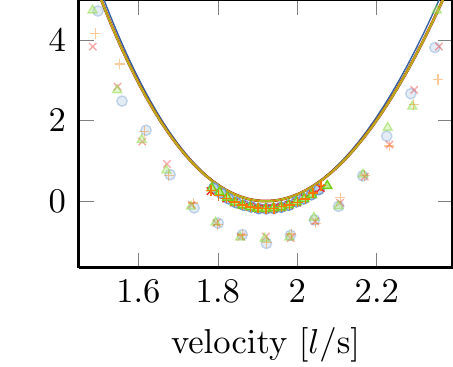}%
  \includegraphics{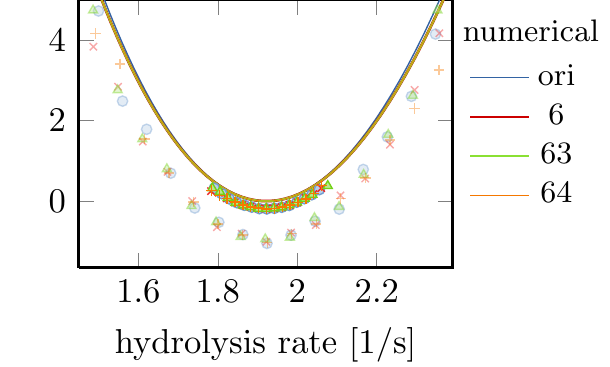}%
  \vspace{-3mm}
  \caption{(color online) Simulation and numerical results for dissipation rate, moving velocity and hydrolysis rate of the kinesin model.
  Data is shown for the original 6-state model (``ori''), a 5-state model with state $6$ reduced (``6'') and two 4-state models with state $6,3$ or $6,4$ reduced (``63'' and ``64''). 
  The rate constants for the original model are taken from Ref.~\cite{Liepelt+Lipowsky2007} describing the data in Ref.~\cite{Carter+Cross2005} for chemical concentrations $c_{ADP}=c_{P}=c_{ATP}= 1\mu$M and stepping size $l\approx 8$nm.  
  The top row shows the sampled pdf for $\tau \approx 1200$s (opaque symbols) and $\tau \approx 120$s (transparent symbols).
  The bins with the width of half an empirical standard deviations are centered around the empirical mean.
  For the simulation we sampled $N = 5000$ trajectories. 
  The bottom row shows convergence of rescaled data (\cf Eq.~(\ref{eq:currentLDP_defn})) to  the rate function $I(s)$ (solid lines).
  }
  \label{fig:fluctuations}
\end{figure*}
To investigate the steady-state fluctuations of the observables we consider stochastic trajectories $\omega=(\omega_0,\omega_1,..,\omega_{N_\omega})$ featuring $N_\omega$ jumps in a prescribed time $\tau_\omega=\tau$.
The time-averaged mean of current observable $O$ along trajectory $\omega$,
\begin{equation}
  j^O_\tau(\omega) = \frac{1}{\tau}\sum_{i=1}^{N_{\omega}}O^{\omega_{i-1}}_{\omega_i},
  \label{eq:epr-for-trajectory}
\end{equation}
is a bounded random variable with the distribution function $f_\tau^{O}$.
For $\tau\to\infty$ it converges weakly and fulfills a large-deviation principle, \ie 
\begin{equation}
  f^{O}_\tau(s) = \exp\left(-\tau I_O(s) + o(\tau) \right),
  \label{eq:currentLDP_defn}
\end{equation}
where $o(\tau)$ stands for a term sublinear in $\tau$.
Further, by the G\"artner-Ellis theorem \cite{Touchette2009}, the large-deviation function $I_O(s)$ is the unique Legendre transform of the scaled cumulant generating function (SCGF)
\begin{equation}
  \zeta(\lambda) = \lim_{\tau \to\infty}\frac{1}{\tau}\log \mathbb E\left[ \exp\left( \lambda \tau j^O_{\tau} \right) \right]
  \label{eq:SCGFdefn}
\end{equation}
where $\mathbb E[\cdot]$ denotes the expectation value on the space of trajectories running for time $\tau$.
The SCGF can be calculated \cite{Touchette2009} as the dominant eigenvalue  of the tilted transition matrix $W_O(\lambda)$ with entries
\begin{equation}
  (W_O)^i_j = w^i_j\exp\left( \lambda O^i_j\right).
  \label{eq:tiltedMatrix}
\end{equation}
To obtain numerical data for the rate function $I_O(s)$ of an observable $O$ one follows the numerical scheme: Calculate $W_O(\lambda)$, determine its largest eigenvalue $\zeta(\lambda)$ and find its Legendre transform with respect to $\lambda$.  For the last step, the algorithm described in Ref.~\cite{Lucet1997} was used.

Fig.\,\ref{fig:4stateFluctuations} shows such numerical results for different physical currents of the enzyme model (Fig.~\ref{fig:bridge-cg}b).
Unlike the na\"ive choice for the rates (dashed lines), our CG method (dotted lines) preserves steady-state averages and fluctuations of the original model (solid lines) to a very high degree.
Bounds of the deviation can be obtained from inequalities for Perron-Frobenius eigenvalues \cite{Altaner+Vollmer2012long}.

\paragraph{Multiple cycles: Kinesin's network of states}

%
Our CG mechanism also captures fluctuations of observables for finite times and in models with multiple cycles.
Figure \ref{fig:reductionkinesin}(c) shows kinesin's network of states \cite{Liepelt+Lipowsky2007}.
Using our method (\ref{eq:newrates}) we reduced the bridge states appearing in the diagram.
The result of a successive reduction of states $6$ and $3$ is shown in \ref{fig:reductionkinesin}(d).
Additionally, we analyzed models with only state $6$ reduced and both states $6$ and $4$ reduced.
Figure \ref{fig:fluctuations} shows the results of simulations, and the convergence to the large-deviation rate function $I(s)$ for the total dissipation rate (entropy production in the medium), the steady-state velocity and the hydrolysis rate of the kinesin model.
Already for finite times the agreement between the original and the reduced models is extremely good.
The rate function $I(s)$ for the different models agree extremely well to the level of being indistinguishable in vicinity of the average value.
Only in the far tails one can observe that the result is not exact.

\paragraph{Discussion}
In this Letter we presented a new method to simplify stochastic dynamics on finite state spaces.
A coarse-graining method that preserves the connection with both the underlying microscopic dynamics and the macroscopic thermodynamics was constructed.
Here we considered bridge states, but the same ideas apply to tree-like subgraphs.
Two biochemical examples where considered:
A generic single-cycle model for enzymatic catalysis and a well-established multi-cycle model for the molecular motor kinesin.
The reduction using the new paradigm preserves fluctuations of current observables in great detail.
Future work will focus on coarse-graining that includes changes of the cycle topology.

\acknowledgments{The authors are indebted to L. Rondoni, A. Puglisi,
  S. Herminghaus and H. Touchette for many illuminating discussions.}


\bibliography{/home/baltaner/bib/noneq/noneq,/home/baltaner/bib/misc/misc}

\begin{thebibliography}{21}%
\makeatletter
\providecommand \@ifxundefined [1]{%
 \@ifx{#1\undefined}
}%
\providecommand \@ifnum [1]{%
 \ifnum #1\expandafter \@firstoftwo
 \else \expandafter \@secondoftwo
 \fi
}%
\providecommand \@ifx [1]{%
 \ifx #1\expandafter \@firstoftwo
 \else \expandafter \@secondoftwo
 \fi
}%
\providecommand \natexlab [1]{#1}%
\providecommand \enquote  [1]{``#1''}%
\providecommand \bibnamefont  [1]{#1}%
\providecommand \bibfnamefont [1]{#1}%
\providecommand \citenamefont [1]{#1}%
\providecommand \href@noop [0]{\@secondoftwo}%
\providecommand \href [0]{\begingroup \@sanitize@url \@href}%
\providecommand \@href[1]{\@@startlink{#1}\@@href}%
\providecommand \@@href[1]{\endgroup#1\@@endlink}%
\providecommand \@sanitize@url [0]{\catcode `\\12\catcode `\$12\catcode
  `\&12\catcode `\#12\catcode `\^12\catcode `\_12\catcode `\%12\relax}%
\providecommand \@@startlink[1]{}%
\providecommand \@@endlink[0]{}%
\providecommand \url  [0]{\begingroup\@sanitize@url \@url }%
\providecommand \@url [1]{\endgroup\@href {#1}{\urlprefix }}%
\providecommand \urlprefix  [0]{URL }%
\providecommand \Eprint [0]{\href }%
\providecommand \doibase [0]{http://dx.doi.org/}%
\providecommand \selectlanguage [0]{\@gobble}%
\providecommand \bibinfo  [0]{\@secondoftwo}%
\providecommand \bibfield  [0]{\@secondoftwo}%
\providecommand \translation [1]{[#1]}%
\providecommand \BibitemOpen [0]{}%
\providecommand \bibitemStop [0]{}%
\providecommand \bibitemNoStop [0]{.\EOS\space}%
\providecommand \EOS [0]{\spacefactor3000\relax}%
\providecommand \BibitemShut  [1]{\csname bibitem#1\endcsname}%
\let\auto@bib@innerbib\@empty
\bibitem [{\citenamefont {Esposito}\ and\ \citenamefont {Van~den
  Broeck}(2010)}]{Esposito+vdBroeck2010}%
  \BibitemOpen
  \bibfield  {author} {\bibinfo {author} {\bibfnamefont {M.}~\bibnamefont
  {Esposito}}\ and\ \bibinfo {author} {\bibfnamefont {C.}~\bibnamefont {Van~den
  Broeck}},\ }\href {\doibase 10.1103/PhysRevE.82.011143} {\bibfield  {journal}
  {\bibinfo  {journal} {Phys. Rev. E}\ }\textbf {\bibinfo {volume} {82}},\
  \bibinfo {pages} {011143} (\bibinfo {year} {2010})}\BibitemShut {NoStop}%
\bibitem [{\citenamefont {Liepelt}\ and\ \citenamefont
  {Lipowsky}(2007)}]{Liepelt+Lipowsky2007}%
  \BibitemOpen
  \bibfield  {author} {\bibinfo {author} {\bibfnamefont {S.}~\bibnamefont
  {Liepelt}}\ and\ \bibinfo {author} {\bibfnamefont {R.}~\bibnamefont
  {Lipowsky}},\ }\href@noop {} {\bibfield  {journal} {\bibinfo  {journal}
  {Phys. Rev. Lett.}\ }\textbf {\bibinfo {volume} {98}},\ \bibinfo {pages}
  {258102} (\bibinfo {year} {2007})}\BibitemShut {NoStop}%
\bibitem [{\citenamefont {Seifert}(2005{\natexlab{a}})}]{Seifert2005}%
  \BibitemOpen
  \bibfield  {author} {\bibinfo {author} {\bibfnamefont {U.}~\bibnamefont
  {Seifert}},\ }\href@noop {} {\bibfield  {journal} {\bibinfo  {journal} {Phys.
  Rev. Lett.}\ }\textbf {\bibinfo {volume} {95}},\ \bibinfo {pages} {40602}
  (\bibinfo {year} {2005}{\natexlab{a}})}\BibitemShut {NoStop}%
\bibitem [{\citenamefont {Seifert}(2005{\natexlab{b}})}]{Seifert2005epl}%
  \BibitemOpen
  \bibfield  {author} {\bibinfo {author} {\bibfnamefont {U.}~\bibnamefont
  {Seifert}},\ }\href@noop {} {\bibfield  {journal} {\bibinfo  {journal}
  {Europhys. Lett.}\ }\textbf {\bibinfo {volume} {70}},\ \bibinfo {pages} {36}
  (\bibinfo {year} {2005}{\natexlab{b}})}\BibitemShut {NoStop}%
\bibitem [{\citenamefont {Faggionato}\ and\ \citenamefont
  {Di~Pietro}(2011)}]{Faggionato+dPietro2011}%
  \BibitemOpen
  \bibfield  {author} {\bibinfo {author} {\bibfnamefont {A.}~\bibnamefont
  {Faggionato}}\ and\ \bibinfo {author} {\bibfnamefont {D.}~\bibnamefont
  {Di~Pietro}},\ }\href@noop {} {\bibfield  {journal} {\bibinfo  {journal} {J.
  Stat. Phys.}\ }\textbf {\bibinfo {volume} {143}},\ \bibinfo {pages} {11}
  (\bibinfo {year} {2011})}\BibitemShut {NoStop}%
\bibitem [{\citenamefont {Hill}(1966)}]{Hill1966}%
  \BibitemOpen
  \bibfield  {author} {\bibinfo {author} {\bibfnamefont {T.}~\bibnamefont
  {Hill}},\ }\href@noop {} {\bibfield  {journal} {\bibinfo  {journal} {J.
  Theor. Biol.}\ }\textbf {\bibinfo {volume} {10}},\ \bibinfo {pages} {442}
  (\bibinfo {year} {1966})}\BibitemShut {NoStop}%
\bibitem [{\citenamefont {Hill}(1977)}]{Hill1977}%
  \BibitemOpen
  \bibfield  {author} {\bibinfo {author} {\bibfnamefont {T.}~\bibnamefont
  {Hill}},\ }\href@noop {} {\emph {\bibinfo {title} {Free energy transduction
  in biology}}}\ (\bibinfo  {publisher} {Academic Press New York},\ \bibinfo
  {year} {1977})\BibitemShut {NoStop}%
\bibitem [{\citenamefont {Hallatschek}(2010)}]{Hallatschek2010}%
  \BibitemOpen
  \bibfield  {author} {\bibinfo {author} {\bibfnamefont {O.}~\bibnamefont
  {Hallatschek}},\ }\href@noop {} {\bibfield  {journal} {\bibinfo  {journal}
  {P.N.A.S.}\ }\textbf {\bibinfo {volume} {108}},\ \bibinfo {pages} {1783}
  (\bibinfo {year} {2010})}\BibitemShut {NoStop}%
\bibitem [{\citenamefont {Schnakenberg}(1976)}]{Schnakenberg1976}%
  \BibitemOpen
  \bibfield  {author} {\bibinfo {author} {\bibfnamefont {J.}~\bibnamefont
  {Schnakenberg}},\ }\href@noop {} {\bibfield  {journal} {\bibinfo  {journal}
  {Rev. Mod. Phys.}\ }\textbf {\bibinfo {volume} {48}},\ \bibinfo {pages} {571}
  (\bibinfo {year} {1976})}\BibitemShut {NoStop}%
\bibitem [{\citenamefont {Kalpazidou}(2006)}]{Kalpazidou2006}%
  \BibitemOpen
  \bibfield  {author} {\bibinfo {author} {\bibfnamefont {S.}~\bibnamefont
  {Kalpazidou}},\ }\href@noop {} {\emph {\bibinfo {title} {Cycle
  representations of {M}arkov processes}}},\ Vol.~\bibinfo {volume} {28}\
  (\bibinfo  {publisher} {Springer, Berlin},\ \bibinfo {year}
  {2006})\BibitemShut {NoStop}%
\bibitem [{\citenamefont {Jiang}\ \emph {et~al.}(2004)\citenamefont {Jiang},
  \citenamefont {Qian},\ and\ \citenamefont {Qian}}]{Jiang_etal2004}%
  \BibitemOpen
  \bibfield  {author} {\bibinfo {author} {\bibfnamefont {D.}~\bibnamefont
  {Jiang}}, \bibinfo {author} {\bibfnamefont {M.}~\bibnamefont {Qian}}, \ and\
  \bibinfo {author} {\bibfnamefont {M.-P.}\ \bibnamefont {Qian}},\ }\href@noop
  {} {\emph {\bibinfo {title} {Mathematical theory of nonequilibrium steady
  states}}}\ (\bibinfo  {publisher} {Springer, Berlin Heidelberg},\ \bibinfo
  {year} {2004})\BibitemShut {NoStop}%
\bibitem [{\citenamefont {Puglisi}\ \emph {et~al.}(2010)\citenamefont
  {Puglisi}, \citenamefont {Pigolotti}, \citenamefont {Rondoni},\ and\
  \citenamefont {Vulpiani}}]{Puglisi_etal2010}%
  \BibitemOpen
  \bibfield  {author} {\bibinfo {author} {\bibfnamefont {A.}~\bibnamefont
  {Puglisi}}, \bibinfo {author} {\bibfnamefont {S.}~\bibnamefont {Pigolotti}},
  \bibinfo {author} {\bibfnamefont {L.}~\bibnamefont {Rondoni}}, \ and\
  \bibinfo {author} {\bibfnamefont {A.}~\bibnamefont {Vulpiani}},\ }\href@noop
  {} {\bibfield  {journal} {\bibinfo  {journal} {J. Stat. Mech.}\ }\textbf
  {\bibinfo {volume} {2010}},\ \bibinfo {pages} {P05015} (\bibinfo {year}
  {2010})}\BibitemShut {NoStop}%
\bibitem [{\citenamefont {Yildiz}\ \emph {et~al.}(2004)\citenamefont {Yildiz},
  \citenamefont {Tomishige}, \citenamefont {Vale},\ and\ \citenamefont
  {Selvin}}]{Yildiz_etal2004}%
  \BibitemOpen
  \bibfield  {author} {\bibinfo {author} {\bibfnamefont {A.}~\bibnamefont
  {Yildiz}}, \bibinfo {author} {\bibfnamefont {M.}~\bibnamefont {Tomishige}},
  \bibinfo {author} {\bibfnamefont {R.~D.}\ \bibnamefont {Vale}}, \ and\
  \bibinfo {author} {\bibfnamefont {P.~R.}\ \bibnamefont {Selvin}},\
  }\href@noop {} {\bibfield  {journal} {\bibinfo  {journal} {Science}\ }\textbf
  {\bibinfo {volume} {303}},\ \bibinfo {pages} {676} (\bibinfo {year}
  {2004})}\BibitemShut {NoStop}%
\bibitem [{\citenamefont {Carter}\ and\ \citenamefont
  {Cross}(2005)}]{Carter+Cross2005}%
  \BibitemOpen
  \bibfield  {author} {\bibinfo {author} {\bibfnamefont {N.}~\bibnamefont
  {Carter}}\ and\ \bibinfo {author} {\bibfnamefont {R.}~\bibnamefont {Cross}},\
  }\href@noop {} {\bibfield  {journal} {\bibinfo  {journal} {Nature}\ }\textbf
  {\bibinfo {volume} {435}},\ \bibinfo {pages} {308} (\bibinfo {year}
  {2005})}\BibitemShut {NoStop}%
\bibitem [{\citenamefont {Lipowsky}\ and\ \citenamefont
  {Liepelt}(2008)}]{Lipowsky+Liepelt2008}%
  \BibitemOpen
  \bibfield  {author} {\bibinfo {author} {\bibfnamefont {R.}~\bibnamefont
  {Lipowsky}}\ and\ \bibinfo {author} {\bibfnamefont {S.}~\bibnamefont
  {Liepelt}},\ }\href {http://dx.doi.org/10.1007/s10955-007-9425-7} {\bibfield
  {journal} {\bibinfo  {journal} {J. Stat. Phys.}\ }\textbf {\bibinfo {volume}
  {130}},\ \bibinfo {pages} {39} (\bibinfo {year} {2008})}\BibitemShut
  {NoStop}%
\bibitem [{\citenamefont {Feller}(1968)}]{Feller1968}%
  \BibitemOpen
  \bibfield  {author} {\bibinfo {author} {\bibfnamefont {W.}~\bibnamefont
  {Feller}},\ }\href@noop {} {\emph {\bibinfo {title} {{An Introduction to
  Probability Theory and Its Applications}}}},\ \bibinfo {edition} {3rd}\ ed.,\
  Vol.~\bibinfo {volume} {1}\ (\bibinfo  {publisher} {Wiley},\ \bibinfo {year}
  {1968})\BibitemShut {NoStop}%
\bibitem [{\citenamefont {Seifert}(2011)}]{Seifert2011}%
  \BibitemOpen
  \bibfield  {author} {\bibinfo {author} {\bibfnamefont {U.}~\bibnamefont
  {Seifert}},\ }\href@noop {} {\bibfield  {journal} {\bibinfo  {journal} {The
  European Physical Journal E: Soft Matter and Biological Physics}\ }\textbf
  {\bibinfo {volume} {34}},\ \bibinfo {pages} {1} (\bibinfo {year}
  {2011})}\BibitemShut {NoStop}%
\bibitem [{\citenamefont {Altaner}\ \emph {et~al.}(2012)\citenamefont
  {Altaner}, \citenamefont {Grosskinsky}, \citenamefont {Herminghaus},
  \citenamefont {Katth{\"a}n}, \citenamefont {Timme},\ and\ \citenamefont
  {Vollmer}}]{Altaner_etal2012}%
  \BibitemOpen
  \bibfield  {author} {\bibinfo {author} {\bibfnamefont {B.}~\bibnamefont
  {Altaner}}, \bibinfo {author} {\bibfnamefont {S.}~\bibnamefont
  {Grosskinsky}}, \bibinfo {author} {\bibfnamefont {S.}~\bibnamefont
  {Herminghaus}}, \bibinfo {author} {\bibfnamefont {L.}~\bibnamefont
  {Katth{\"a}n}}, \bibinfo {author} {\bibfnamefont {M.}~\bibnamefont {Timme}},
  \ and\ \bibinfo {author} {\bibfnamefont {J.}~\bibnamefont {Vollmer}},\
  }\href@noop {} {\bibfield  {journal} {\bibinfo  {journal} {Physical Review
  E}\ }\textbf {\bibinfo {volume} {85}},\ \bibinfo {pages} {041133} (\bibinfo
  {year} {2012})}\BibitemShut {NoStop}%
\bibitem [{\citenamefont {Altaner}\ and\ \citenamefont
  {Vollmer}(shed)}]{Altaner+Vollmer2012long}%
  \BibitemOpen
  \bibfield  {author} {\bibinfo {author} {\bibfnamefont {B.}~\bibnamefont
  {Altaner}}\ and\ \bibinfo {author} {\bibfnamefont {J.}~\bibnamefont
  {Vollmer}},\ }\href@noop {} {\  (\bibinfo {year} {unpublished})}\BibitemShut
  {NoStop}%
\bibitem [{\citenamefont {Touchette}(2009)}]{Touchette2009}%
  \BibitemOpen
  \bibfield  {author} {\bibinfo {author} {\bibfnamefont {H.}~\bibnamefont
  {Touchette}},\ }\href@noop {} {\bibfield  {journal} {\bibinfo  {journal}
  {Physics Reports}\ }\textbf {\bibinfo {volume} {478}},\ \bibinfo {pages} {1}
  (\bibinfo {year} {2009})}\BibitemShut {NoStop}%
\bibitem [{\citenamefont {Lucet}(1997)}]{Lucet1997}%
  \BibitemOpen
  \bibfield  {author} {\bibinfo {author} {\bibfnamefont {Y.}~\bibnamefont
  {Lucet}},\ }\href@noop {} {\bibfield  {journal} {\bibinfo  {journal}
  {Numerical Algorithms}\ }\textbf {\bibinfo {volume} {16}},\ \bibinfo {pages}
  {171} (\bibinfo {year} {1997})}\BibitemShut {NoStop}%
\end{thebibliography}%

\end{document}